\documentclass[aps,prb,twocolumn,showpacs]{revtex4}
\usepackage{amsmath} 
\usepackage{graphicx} 
 
\newcommand{\lsim}{\lesssim} 
\newcommand{\gsim}{\gtrsim} 

\newcommand{\ket}[1]{|#1\rangle}
\newcommand{\bra}[1]{\langle#1|}
\newcommand{\tr}{\mathrm{Tr}}
\newcommand{\spol}{\tfrac{1}{2}}
\newcommand{\pol}{\frac{1}{2}}
\newcommand{\up}{\uparrow}
\newcommand{\dow}{\downarrow}
\newcommand{\ok}[1]{\left( #1 \right)}
\newcommand{\braket}[2]{\langle #1 | #2 \rangle}

\begin{document} 
 
\title{Enhanced Conductance Through Side-Coupled Double Quantum Dots} 
 
\author{R. \v{Z}itko$^1$} 
\author{J. Bon\v{c}a$^{1,2}$} 
 
\affiliation{$^1$J. Stefan Institute, SI-1000 Ljubljana, 
and $^2$Department  of Physics, 
FMF, University of Ljubljana, SI-1000 Ljubljana, Slovenia}
 
\date{\today} 
 
\begin{abstract}
Conductance, on-site and inter-site charge fluctuations and 
spin correlations in the system of two side-coupled quantum dots 
are calculated using the Wilson's numerical renormalization group (NRG) 
technique. We also show spectral density calculated using the density-matrix NRG,
which for some parameter ranges remedies inconsistencies of the conventional
approach.
By changing the gate voltage and the inter-dot tunneling rate, 
the system can be tuned to a non-conducting spin-singlet state, 
the usual Kondo regime with odd number of electrons occupying the dots,
the two-stage Kondo regime with two electrons, or a valence-fluctuating state 
associated with a Fano resonance.
Analytical expressions for the width of the Kondo regime
and the Kondo temperature are given. We also study the effect of unequal
gate voltages and the stability of the two-stage Kondo effect with
respect to such perturbations.
\end{abstract} 
 
 
\maketitle 
 
\section{Introduction}

The advances in micro-fabrication have enabled studies of transport
through single as well as coupled quantum dots, where at very low
temperatures Kondo physics and magnetic interactions play an important
role. A double-dot system represents the simplest possible
generalization of a single-dot system which has been extensively
studied in the past. Recent experiments demonstrate that an
extraordinary control over the physical properties of double dots can
be achieved \cite{dqd-expr1,dqd-expr2,dqd-expr3,pcdqd}, which enables
direct experimental investigations of the competition between the
Kondo effect and the exchange interaction between localized moments on
the dots. One manifestation of this competition is a two stage Kondo
effect that has recently been predicted in multilevel quantum dot
systems with explicit exchange interaction coupled to one or two
conduction channels \cite{qptmultilevel, hofstetter2004}. 
Experimentally, it manifests itself as a sharp drop in the conductance 
vs. gate voltage $G(V_G)$ \cite{two_stage} or as non-monotonic dependence of 
the differential conductance vs. drain-source voltage $dI/dV_{ds}(V_{ds})$ 
\cite{granger2005}.

Fano resonances, which occur due to interference when a 
discrete level is weakly coupled to a continuous band, 
were recently observed in experiments on rings with embedded 
quantum dots \cite{ringfano} and quantum wires with side-coupled 
dots \cite{scfano}.  
The interplay between Fano and Kondo resonance was investigated using 
equation of motion  \cite{bulka1,bulka2} 
and slave boson techniques \cite{sidedouble}. 
 
\begin{figure}[htb]
\centering
\includegraphics[totalheight=2cm]{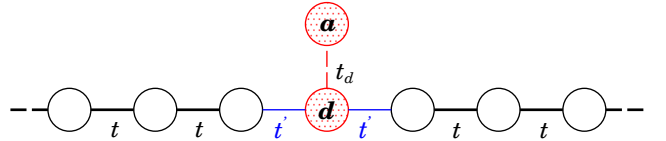}
\caption{Side-coupled configuration of quantum dots} 
\label{figa1}
\end{figure}

In this work we study a double quantum dot (DQD) in a side-coupled
configuration (Fig. \ref{figa1}), connected to a single
conduction-electron channel. Systems of this type were studied
previously using non-crossing approximation \cite{suppression},
embedding technique \cite{topology} and slave-boson mean field theory
\cite{kang,sidedouble}. 
Numerical renormalization group (NRG) calculations were also performed
recently \cite{corn}, where only narrow regimes of enhanced conductance
were found at low temperatures.
We will show that when the intra-dot overlap is large, wide regimes of
enhanced conductance as a function of gate-voltage exist at low
temperatures due to the Kondo effect, separated by regimes where
localized spins on DQD are antiferromagnetically (AFM) coupled.  Kondo
temperatures $T_K$ follow a prediction based on Schrieffer-Wolff transformation 
and poor-man's scaling. In the limit when the dot $a$ is
only weakly coupled, the system enters the "two stage" Kondo regime
\cite{vojta,corn}, where we again find a wide regime of enhanced
conductivity under the condition that the high- and the low- Kondo
temperatures ($T_K$ and $T_K^0$ respectively) are well separated and
the temperature of the system $T$ is in the interval $T_K^0 \ll T \ll
T_K$.

\section{Model and method}
 
The Hamiltonian that we study reads 
\begin{equation} 
\begin{split}
H &=  
\delta_d (n_d-1) + \delta_a (n_a-1)
- t_d \sum_\sigma \left( d^\dag_\sigma a_\sigma + a^\dag_\sigma d_\sigma 
\right) \\
&+ \frac{U_d}{2} (n_d-1)^2 + \frac{U_a}{2} (n_a-1)^2 \\ 
&+ \sum_{k\sigma} \epsilon_k c^\dag_{k\sigma} c_{k\sigma} + 
\sum_{k\sigma} V_d(k) \left( c^\dag_{k\sigma} d_{\sigma} + 
d^\dag_\sigma c_{k\sigma} \right), \label{ham}
\end{split}
\end{equation} 
where $n_d=\sum_\sigma d^\dag_\sigma d_\sigma$ and $n_a=\sum_\sigma
a^\dag_\sigma a_\sigma$. Operators $d^\dag_\sigma$ and $a^\dag_\sigma$
are creation operators for an electron with spin $\sigma$ on site $d$
or $a$.  On-site energies of the dots are defined by
$\epsilon=\delta-U/2$. 
For simplicity, we choose the on-site
energies and Coulomb interactions to be equal on both dots,
$\delta_a=\delta_d=\delta$ and $U_a=U_d=U$.
Coupling between the dots is described by the inter-dot tunnel
coupling $t_d$. Dot $d$ couples to both leads with
equal hopping $t'$.  
As it couples only to symmetric combinations of the
states from the left and the right lead, we have used a unitary
transformation
\cite{glazmanraikh} to describe the system as a variety of the
single-channel, two-impurity Anderson model.
Operator $c_{k\sigma}^\dag$ creates a conduction band electron with
momentum $k$, spin $\sigma$ and energy $\epsilon_k=-D \cos{k}$, where
$D=2t$ is the half-bandwidth. 
The momentum-dependent hybridization function is
$V_d(k)=-(2/\sqrt{N+1})\, t' \sin{k}$, where $N$ in the normalization
factor is the number of conduction band states.

We use Meir-Wingreen's formula for conductance in the case of 
proportionate coupling \cite{meirwingreen} which
is known to apply under very general conditions
(for example, the system need not be in a Fermi-liquid ground state)
with spectral functions obtained using the NRG technique 
\cite{wilson, magnetocosti, sia1, hofstetter}.
At zero temperature, the conductance is
\begin{equation} 
G=G_0 \pi \Gamma \rho_d(0), 
\end{equation} 
where $G_0=2e^2/h$, $\rho_d(\omega)$ is the local density of states of 
electrons on site $d$ and $\Gamma/D=(t'/t)^2$.
 
The NRG technique consists of logarithmic discretization
of the conduction band, mapping onto a one-dimensional chain
with exponentially decreasing hopping constants, 
and iterative diagonalization of the resulting Hamiltonian \cite{wilson}. 
Only  low-energy part of the spectrum is kept after each iteration step;
in our calculations we kept 1200 states, not counting spin 
degeneracies, using discretization parameter $\Lambda=1.5$.

\section{Strong inter-dot coupling}

\subsection{Conductance and correlation functions}

\begin{figure}[htbp] 
\includegraphics[width=9cm,angle=-90]{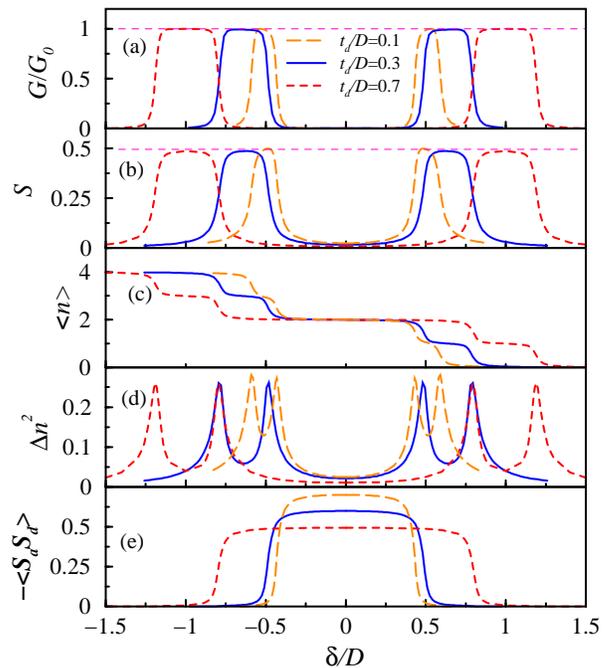} 
\caption{Conductance and correlation functions of DQD vs. $\delta$. Besides 
different values of $t_d$, indicated in the figure, other parameters
of the model are $\Gamma/D=0.03$ and $U/D=1$. 
Temperature is $T/D=10^{-9}$, which for all parameters used corresponds
to a zero temperature limit. In particular, in the Kondo plateaux $T \ll T_K$
for all $\delta$.
}
\label{figa2}
\end{figure} 

In Fig.~\ref{figa2}a we present conductance through a double 
quantum dot at different values of intra-dot couplings vs. $\delta$. 
Due to formation of Kondo correlations, conductance is enhanced,
reaching the unitary limit in a wide range of $\delta$. 

To better understand multidot problems in the case of strong inter-dot coupling, 
it is helpful to exactly diagonalize the part of the Hamiltonian that 
describes the dots and rewrite the entire Hamiltonian (Eq.~\ref{ham}) in a form similar to 
the ionic model \cite{hewson}:
\begin{equation}
H_0=\sum_{k,\sigma} \epsilon_k c^\dag_{k,\sigma} c_{k,\sigma}
+\sum_{\alpha} E(\alpha) \ket{\alpha} \bra{\alpha}
\end{equation}
\begin{equation}
H_1=\sum_{k,\sigma,\alpha,\beta}
t_{k\sigma|\beta\to\alpha}
\ \ket{\alpha} \bra{\beta}
\ c_{k,\sigma}
+t^*_{k\sigma|\beta\to\alpha} c^\dag_{k\sigma}
\ \ket{\beta} \bra{\alpha}.
\end{equation}
where multi-indeces $\alpha$ and $\beta$ stand for quantum numbers $(Q,S,S_z,r)$.
Here $Q$ is the charge number $Q=(n_a-1)+(n_d-1)$, $S, S_z$ are the spin and
its component, while $r$ numbers different states with the same $Q,S,S_z$ quantum
numbers. In the absence of the magnetic field, $S_z$ is irrelevant and 
will be omitted from now on.
Each $(Q,S,r)$ multiplet is then $2S+1$-fold degenerate.
Finally, the effective hopping coefficients
\begin{equation}
t_{k\sigma|\beta\to\alpha} = V_d(k) \bra{\alpha} d_\sigma^\dag \ket{\beta}
\end{equation}
correspond to electrons hopping from the conduction band to the dots.

\begin{figure}[htbp]
\includegraphics[width=8.5cm]{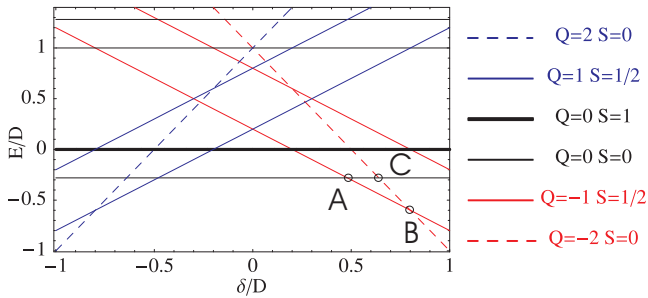}
\caption{Eigenvalue diagram for isolated DQD system. 
The diagram is symmetric, since for $\delta \to -\delta$,
$E(Q,S,r) \to E(-Q,S,r)$.
Points $A$ and $B$ correspond to valence-fluctuation regions 
when the charge on the dot changes, while point $C$ corresponds 
to the center of the Kondo regime, when the Kondo temperature is 
the lowest.
Parameters are $U/D=1$ and $t_d/D=0.3$.
}
\label{figa3}
\end{figure}

The eigenvalue diagram in Fig.~\ref{figa3} represents the gate-voltage 
dependence of the multiplet energies $E(Q,S,r)$. From this diagram 
we can read off the ground state and the excited states for 
each parameter $\delta$. 

\begin{figure}[htbp] 
\includegraphics[width=8cm]{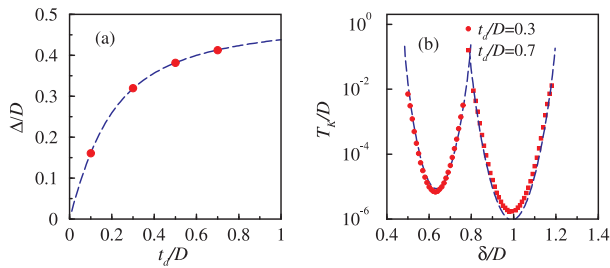} 
\caption{a) The width of conductance peak $\Delta$ vs. $t_d$ as
obtained from NRG calculations (full circles), compared with the
analytical result as given in the text (dashed line).  b) Kondo
temperatures $T_K$ vs. $\delta$ as obtained from the widths of Kondo
peaks (full circles). Analytical estimate, Eq.~\ref{tk}, is shown
using dashed lines.
The rest of parameters are identical to those in 
Fig.~\ref{figa2}.}
\label{figa4}
\end{figure}

Regimes of enhanced
conductance appear in the intervals approximately given by
$\delta_1<|\delta|< \delta_2$, where
$\delta_1= t_d (2\sqrt{1+(U/4t_d)^2}-1)$ and
$\delta_2=(U/2+t_d)$. 
These estimates are obtained from the lowest
energies of states with zero, one and two electrons on the isolated double
quantum dot:
\begin{equation}
\begin{split}
E(-2,0,0) &= U - 2\delta, \\
E(-1,\spol,0) &= U/2-\delta-t_d, \\
E(0,0,0) &= -2t_d\sqrt{1+(U/4t_d)^2}+U/2.
\end{split}
\end{equation}
The widths of conductance peaks (measured at $G/G_0=1/2$) 
are therefore approximately given
by 
\begin{equation}
\Delta= U/2 + 2t_d(1 - \sqrt{1+(U/4t_d)^2}).
\end{equation}
Note that in the limit of large $t_d$, $\Delta\simeq U/2$, 
and in the limit of large
$U$, $\Delta\simeq 2t_d$. As it will become apparent later in this
paper, this naive estimate fails when $t_d \to 0$. 
Comparison of conductance-peak widths with the
analytical estimate $\Delta$ is shown in Fig.~\ref{figa4}a.

Next, we focus on various correlation functions. In Fig.~\ref{figa2}b
we show $S$, calculated from expectation value 
$\langle {\bf S}_\mathrm{tot}^2 \rangle=S(S+1)$, 
where ${\bf S}_\mathrm{tot} = {\bf S}_a+{\bf S}_d$ is the total spin operator. 
$S$ reaches value $1/2$ in the regime where $G/G_0=1$. Enhanced
conductance is thus followed by the local moment formation, indicative
of the Kondo effect. This is further supported by the average
double-dot occupancy $\langle n\rangle $, where $n=n_a+n_d$, which in
the regime of enhanced conductivity approaches odd-integer values,
{i.e.} $\langle n
\rangle=1$ and $3$ (see Fig.~\ref{figa2}c). Transitions between regimes
of nearly integer occupancies are rather sharp; they are visible as
regions of enhanced charge fluctuations measured by $\Delta n^2 =
\langle n^2\rangle - \langle n\rangle^2$, as shown in
Fig.~\ref{figa2}d. Finally, we show in Fig.~\ref{figa2}e spin-spin
correlation function $\langle {\bf S}_a \cdot {\bf S}_d \rangle$. Its value
is negative between two separated Kondo regimes where conductance
approaches zero, {\it i.e.} for $-\delta_1<\delta<\delta_1$, otherwise
it is zero. This regime further coincides with $\langle n\rangle\sim
2$. Each dot thus becomes nearly singly occupied and spins on the two
dots form a local singlet due to effective exchange coupling
$J_{\mathrm{eff}} \approx 4t_d^2/U$. Surprisingly, the absolute value of
$\langle{ {\bf S}_a \cdot {\bf S}_d}\rangle$ increases and it nearly reaches
its limiting value, {\it i.e.}  $\langle {\bf S}_a \cdot {\bf S}_d \rangle =
-3/4$, as the intra-dot hopping $t_d$, and with it $J_{\mathrm{eff}}$,
decrease. This seeming contradiction is resolved by recognizing that
$J_{\mathrm{eff}}$ represents the exchange interaction based on the
effective Heisenberg coupling between localized spins on DQD only in
the limit when $t_d/U\to 0$. Smaller absolute values of $\langle{ {\bf
S}_a \cdot {\bf S}_d}\rangle$ are thus due to a small amount of  double 
occupancy at larger values of $t_d/U$.

In Fig.~\ref{figa4}b we present Kondo temperatures
$T_K$ vs. $\delta$ extracted from the widths of Kondo
peaks. Numerical results in the regime where $\langle n \rangle\sim 1$
and 3 fit the analytical expression obtained using the Schrieffer-Wolff
transformation that leads to an effective single $S=1/2$ Kondo model.
We obtain
\begin{equation} 
T_K=0.182 U \sqrt{\rho_0 J}\exp[-1/\rho_0J] \label{tk} 
\end{equation} 
with 
\begin{eqnarray}
\rho_0 J&=&\frac{2\Gamma}{\pi} \Biggl( \frac{\alpha}{\vert E(-1,\spol,0)
-E(-2,0,0)\vert} \nonumber \\ &+&
\frac{\beta}{\vert E(-1,\spol,0)-E(0,0,0) \vert} \Biggr),
\label{rhoJ}
\end{eqnarray}
where 
\begin{equation} 
\begin{split}
\alpha &= |\bra{-1,\spol,\spol,0} d^\dag_\up \ket{-2,0,0,0}|^2=1/2, \\
\beta &= |\bra{0,0,0,0} d^\dag_\dow \ket{-1,\spol,\spol,0}|^2 \\
&= \frac{
\left(4 t_d + U + \sqrt{16 t_d^2+U^2}\right)^2
}{
8\left(16 t_d^2+U\left(U+\sqrt{16 t_d^2+U^2}\right)\right)
}.
\end{split}
\label{beta}
\end{equation} 
The factor $0.182U$ in Eq.~\ref{tk} is the effective bandwidth. 
The same effective bandwidth was used to obtain $T_K$ of the Anderson 
model in the regime $U<D$ \cite{sia1,hald}.

\subsection{Spectral densities}

\begin{figure}[htbp]
\includegraphics[width=5.5cm]{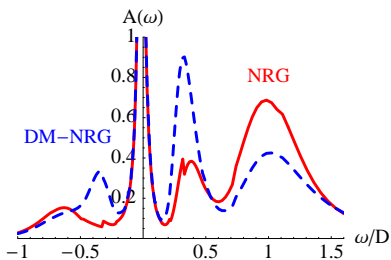}
\caption{Comparison of spectral densities computed by the conventional
NRG and by the DM-NRG for parameters where the conventional NRG approach has
difficulties: $t_d/D=0.3$, $U/D=1$, $\Gamma/D=0.03$, $\delta/D=0.48$. 
Note the discontinuities in the spectral density calculated using the conventional NRG.
At low frequencies both method produce identical results, so we
do not show the Kondo peak which exceeds the vertical scale of the plot.
In this calculations, the effective temperature is zero.
}
\label{figa5}
\end{figure}

We observed that the spectral density calculation using the conventional
NRG approach \cite{costi} fails for our model. The spectral densities manifest
spurious discontinuities and the normalization sum rule is violated for
some choices of model parameters (Fig.~\ref{figa5}). This happens because at
the early stages of the NRG iteration the lowest energy state does
not yet correspond to the true ground state.
We found that we obtain correct results using the density-matrix
NRG technique \cite{hofstetter},
which remedies the shortcomings of the conventional approach.
We succeeded in implementing this technique with eigenstates
defined within subspaces with well defined occupation $Q$ and
total spin $S$, as opposed to well defined occupation and
total spin component $S_z$ (see the Appendix).
The improvement in numerical efficiency is sufficient to 
enable consideration of more complex systems, such as the double 
quantum dot.

\begin{figure}[htbp]
\includegraphics[width=8cm]{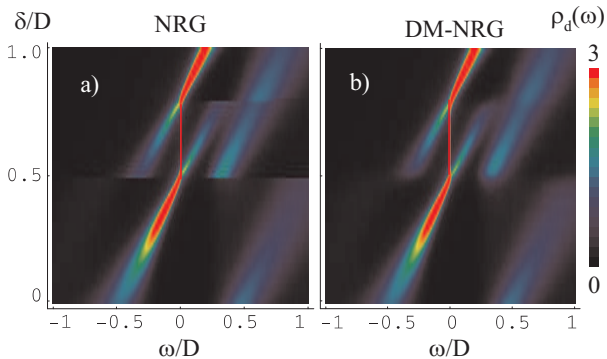}
\caption{Zero-temperature spectral density $\rho_d(\omega)$ sweeps for
$t_d/D=0.3$. a) Spectral density calculated using the conventional
NRG approach. b) Spectral density calculated using
the density-matrix NRG approach.
Note that the vertical line, representing the Kondo
peak, has been artificially broadened. Its true width is given by
$T_K$.
}
\label{figa6}
\end{figure} 

\begin{figure}[htbp]
\includegraphics[width=8cm]{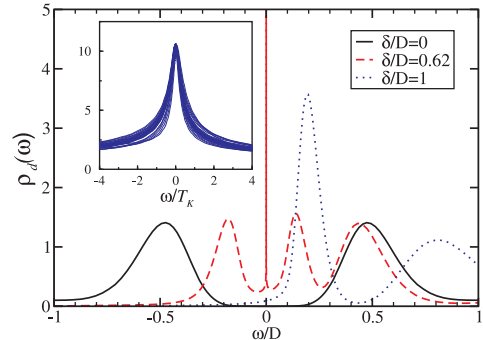}
\caption{Three zero-temperature spectral densities $\rho_d(\omega)$ 
for $t_d/D=0.3$:
at the particle-hole symmetric point $\delta/D=0$, in the Kondo
regime $\delta/D=0.62$ and in the empty-orbital regime $\delta/D=1$.
{\sl Inset}: Scaling of spectral densities $\rho_d(\omega/T_K)$
in the Kondo region $0.52 < \delta < 0.76$. Parameters are as in Fig.~\ref{figa2}.}
\label{figa7}
\end{figure}

In Fig.~\ref{figa6} we present sweeps of $\rho_d(\omega)$ calculated
using both approaches. In vast regions of the plot the
results are in perfect agreement. The differences appear for
those values of $\delta$ where the ground state changes.
Three characteristic spectral density curves calculated using the DM-NRG
are shown in Fig.~\ref{figa7}.

Features in the spectral density sweeps can be easily interpreted
using eigenvalue diagram in Fig.~\ref{figa3}. 
At low temperatures and for constant $\delta$, spectral density 
$A(\omega)$ will be high whenever the energy difference $\Delta E=E_1-E_0$ 
between the ground state $(0)$ and an excited state $(1)$ is equal 
to $+\omega$ (particle excitations, $Q_1=Q_0+1$, $S_1=S_0\pm\spol$) 
or to $-\omega$ (hole excitations, $Q_1=Q_0-1$, $S_1=S_0\pm\spol$).
At $\delta=0$ two broad peaks are seen located symmetrically at
$\omega\sim\pm\delta_1$ (see Fig.~\ref{figa6} and Fig.~\ref{figa7},
at $\delta/D=0$). 
At this point the model is particle-hole symmetric
and therefore $E(Q,S,r)=E(-Q,S,r)$ for all $Q,S,r$. Consequently,
the spectrum is also symmetric, $\rho_d(\omega)=\rho_d(-\omega)$. 
With increasing $\delta$, the
particle excitation energy $E(1,\spol,0)-E(0,0,0)$ increases
and the corresponding peak quickly washes out. The hole excitation
energy $E(-1,\spol,0)-E(0,0,0)$ decreases and the peak gains
weight. 

At $\delta=\delta_1$, $E(0,0,0) = E(-1,\spol,0)$ 
(point $A$ in Fig.~\ref{figa3}) 
and the system enters the Kondo regime: a sharp many-body resonance 
appears which remains pinned at the Fermi level throughout the Kondo
region (see Fig.~\ref{figa6} and Fig.~\ref{figa7}, at $\delta/D=0.62$).
Kondo effect occurs whenever the ground state is a doublet, 
$S=1/2$, and there are excited states with $S'=0,1$, $Q'=Q \pm 1$. 
We recognize such regions by characteristic triangular level crossings in
the eigenvalue diagram, one of which is marked as triangle
$ABC$ in Fig.~\ref{figa3}.
The high-energy peaks at $\omega=E(0,0,0) - E(-1,\spol,0)>0$
and $\omega=E(-1,\spol,0)-E(-2,0,0)<0$ in the spectral density are also
characteristic: they correspond to particle and hole excitations 
that are at the heart of the Kondo effect.

In the case of the DQD we also see additional structure for $\delta_1 < \delta < \delta_2$:
a broad peak at $\omega=E(0,1,0)-E(-1,\spol,0)$ which corresponds to virtual
triplet excitations from the ground state. These excitations could also be taken 
into account in the calculation of the effective exchange interaction, Eq.~\eqref{rhoJ},
however due to their high energy, they only lead to an exponentially small 
renormalization of the Kondo temperature, which may be neglected.

In the inset of Fig.~\ref{figa7} we show scaling of Kondo peaks vs. $\omega/T_K$. 
In the case of perfect scaling, all curves should exactly overlap. However,
Kondo temperatures of different peaks differ by almost four orders of magnitudes, 
as seen in Fig.~\ref{figa4}b. Moreover, Kondo peaks become asymmetric near the
edges of the Kondo region, i.e. for $\delta \gsim \delta_1$ and $\delta \lsim \delta_2$.
Note also that for each point in Fig.~\ref{figa4}b there
is a respective spectral density presented in the inset of Fig.~\ref{figa7}.

\subsection{Effect of the on-site-energy splitting}
\label{splitting1}

In real double quantum dot systems, it is difficult to produce dots and electrodes 
with identical properties. In particular, it is not easy to achieve
equal on-site energies $\delta_d=\delta_a=\delta$ for a wide range of parameters 
$\delta$. Therefore it is necessary to study the robustness of the physical
regimes with respect to the splitting of the on-site energies.

We now generalize slightly our model to allow for unequal on-site energies
by introducing a new parameter, $\kappa$, so that
on-site energies are $\delta_d=\delta+\kappa$ and $\delta_a=\delta-\kappa$.
The DQD part of the Hamiltonian is now
\begin{equation}
\begin{split}
H_0 &= (\delta+\kappa) (n_d-1) + (\delta-\kappa) (n_a-1) \\
&- t_d \sum_\sigma \left( d^\dag_\sigma a_\sigma + a^\dag_\sigma d_\sigma 
\right) \\
&+ \frac{U}{2} (n_d-1)^2 + \frac{U}{2} (n_a-1)^2
\end{split}
\end{equation}
%
Only $\kappa>0$ has to be considered, because $H_0(\delta,-\kappa) \to
H_0(-\delta,\kappa)$ when particle-hole transformation $d^\dag_\mu \to
d_\mu, a^\dag_\mu \to -a_\mu$ is performed.

By diagonalization $H_0$ exactly, we find that the effect of
$\kappa$ is to shift states while the general eigenvalue structure is
maintained and is similar to that shown in Fig.~\ref{figa3}. We therefore
still expect to see two Kondo peaks as $\delta$ is swept for a constant
$\kappa$. 

\begin{figure}[htbp]
\includegraphics[width=7cm,angle=-90]{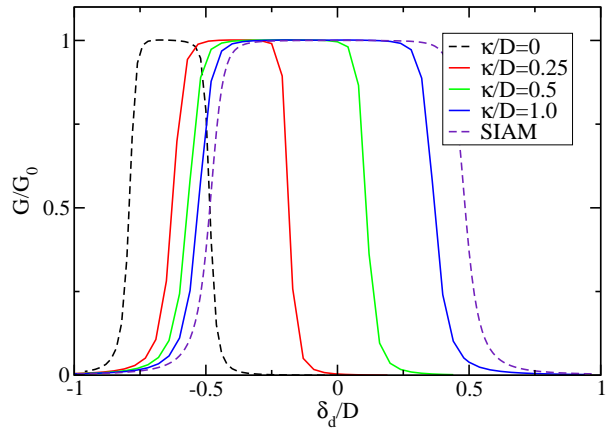}
\caption{Conductance as a function of gate-voltage $\delta_d$
for different gate-voltage asymmetries $\kappa$. 
Label SIAM corresponds to
results for conductance of the corresponding single-impurity Anderson model.
The temperature is $T/D=10^{-9}$.
}
\label{figa8}
\end{figure}

\begin{figure}[htbp]
\includegraphics[width=8cm]{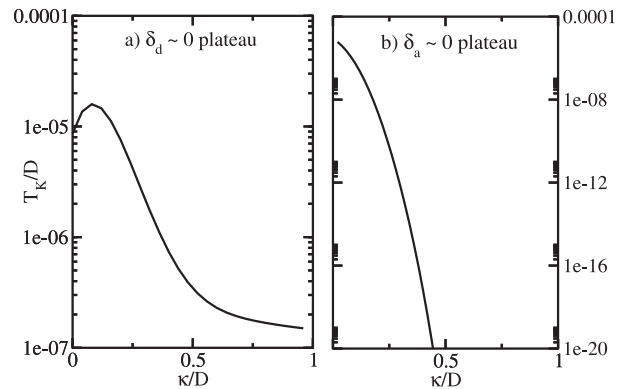}
\caption{Kondo temperature as a function of $\kappa$ for
a) the Kondo plateau at $\delta_d \sim 0$ and b) the
Kondo plateau at $\delta_a \sim 0$.
In both cases we calculate the Kondo temperature in
the middle of the plateau, where $T_K$ is lowest.
Parameters are $U/D=1$, $\Gamma/D=0.03$ and $t_d/D=0.3$.
}
\label{figa9bis}
\end{figure}

In the limit of large $\kappa$ and for $\delta$ such that $\delta_d \sim 0$ 
and $\delta_a \to \pm\infty$, the side-coupled dot becomes irrelevant and 
we recover the familiar single-impurity Anderson model with conduction plateau 
in the region $-U/2 < \delta_d < U/2$ with $T_K$ given by Eq.~\eqref{tk} with
\begin{equation}
\rho_0 J = \frac{2\Gamma}{\pi} \ok{\frac{1}{|\delta_d-U/2|}
+\frac{1}{|\delta_d+U/2}}.
\end{equation}
This scenario is corroborated by NRG calculations (Fig.~\ref{figa8}).
The approach to the limit can be determined using the expressions
Eqs.~\eqref{tk} and \eqref{rhoJ}, with $\kappa$ dependent $\alpha(\kappa)$,
$\beta(\kappa)$ and $E_1(\kappa), E_2(\kappa)$. 
The result is shown in Fig.~\ref{figa9bis}a.
The $\kappa$ dependence is non-monotonic, which can be traced to
non-monotonic behavior of coefficient $\beta(\kappa)$. At
$\kappa/D=1$ the asymptotic value $T_K/D=1.04 \times 10^{-7}$
is nearly reached.

\begin{figure}[htbp]
\includegraphics[width=7cm,angle=-90]{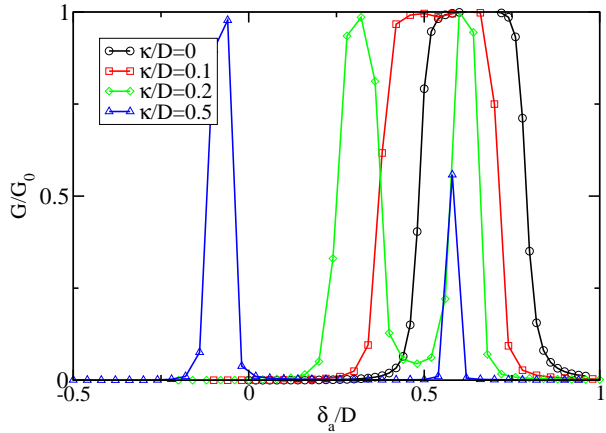}
\caption{Conductance as a function of gate-voltage $\delta_a$
for different gate-voltage asymmetries $\kappa$. 
The temperature is fixed to $T/D=10^{-9}$ for all $\kappa$.
}
\label{figa9}
\end{figure}

In the limit of large $\kappa$, but for $\delta$ such that $\delta_a \sim 0$ and
$\delta_d \to \pm\infty$, the embedded dot will be either 
fully occupied or empty with very little charge fluctuations, while
the side-coupled dot will maintain $S=1/2$ spin moment. 
This moment will be screened, but due to the suppression of the 
virtual electron hopping through the dot $d$ the corresponding
Kondo temperature rapidly drops with increasing $\kappa$ (Fig.~\ref{figa9bis}b).
Keeping the temperature constant and increasing $\kappa$, 
the Kondo plateau in the region $-U/2 < \delta_a < U/2$ 
will therefore quickly evolve into two peaks separated by a Coulomb blockade valley
and for very large $\kappa$ it will not conduct at all.
This prediction can also be verified using NRG (Fig.~\ref{figa9}).

\section{Weak inter-dot coupling}

\subsection{Conductance and correlation functions}

\begin{figure}[htbp] 
\includegraphics[width=6.5cm,angle=-90]{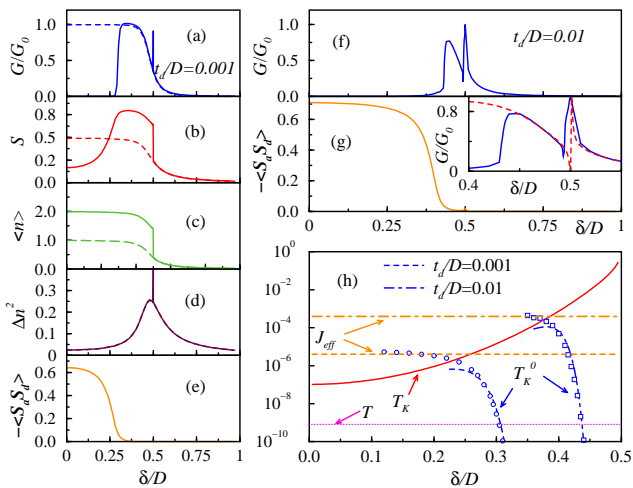} 
\caption{Conductance and correlation functions 
at $t_d/D=0.001$ (a,...,e), $t_d/D=0.01$ (f,and g) and a blow-up of f)
in the inset of g). Dashed lines in a) represent $G/G_0$ and b) $S$ of
a single quantum dot with otherwise identical parameters. Dashed line
in c) represents $\langle n_d\rangle$ of DQD, and finally dashed line
in the inset of g) represents the semi-analytical model described in
the text.  In h) a schematic plot of different temperatures and
interactions is presented as explained in the text. NRG values of the
gap in $\rho_d(\omega)$ at $\omega=0$ and $T\ll T_K^0$ are presented
with open circles and squares for $t_d/D=0.001$ and $0.01$
respectively.  Values of $J_{\mathrm{eff}}$ and analytical results of $T_K^0$
are presented with dashed and dot-dashed lines for $t_d/D=0.001$ and
$0.01$ respectively. For analytical estimates of $T_K^0$ different
values of $\alpha=2.2$ and $1.3$, respectively, were used. Other
parameters of the model are $\Gamma/D=0.03, U/D=1$.}
\label{figa10}
\end{figure}

We now turn to the limit when $t_d\to 0$. Unless otherwise
specified, we choose
the effective temperature $T$ to be finite,
{\it i.e.} $T\sim 10^{-9}D$, since calculations at much lower 
temperatures would be experimentally irrelevant.
In this case one
naively expects to obtain essentially identical conductance as in the
single-dot case.  As $\delta$ decreases below $\delta\sim U/2$,
$G/G_0$ indeed follows result obtained for the single-dot case as
shown in Fig.~\ref{figa10}a. In the case of DQD, however, a sharp Fano
resonance appears at $\delta = U/2$. This resonance coincides with the
sudden jump in $S$, $\langle n\rangle$, as well as with the spike in
$\Delta n^2$, as shown in Figs~\ref{figa10}b,c, and d,
respectively. Fano resonance is a consequence of a sudden charging of
the nearly decoupled dot $a$, as its level $\epsilon$ crosses the
chemical potential of the leads, {\it i.e} at $\epsilon=0$. Meanwhile,
the electron density on the dot $d$ remains a smooth function of
$\delta$, as seen from $\langle n_d\rangle$ in Fig~\ref{figa10}c. With
increasing $t_d$, the width of the resonance increases, as shown for
$t_d/D=0.01$ in Fig.~\ref{figa10}f. For $t_d \gtrsim 0.1$, the resonance merges
with the Kondo plateau and disappears (see Fig.~\ref{figa2}a).
 
The resonance at $\delta \sim U/2$ can be explained using a simple
semi-analytical model.  For $t_d=0$, we write the exact Green's
function of the impurity $d$ at $\omega=0$ as
${\cal G}^0_d=1/(i\Gamma-\Gamma\cot\phi)$
\cite{langreth66}, where $\phi$ is the scattering phase shift for 
single impurity model that we calculate using NRG.  The side-coupled
dot is taken into account using perturbation theory.  We obtain full
Green's function from the Dyson's equation 
\begin{equation}
{\cal G}_d={\cal G}_d^0/(1-t_d^2 {\cal G}_a^0 {\cal G}_d^0).
\end{equation}
For the Fano resonance at $\epsilon = \delta-U/2 \sim 0$,
we keep only the low-energy pole in the Green's function, 
\begin{equation}
{\cal G}_a^0(\omega) \approx \tfrac{z}{\omega-\epsilon+i\eta},
\end{equation}
where $z=\tfrac{1}{2}$ for $\epsilon<0$ and $z=1$ for $\epsilon>0$.
The conductance is
\begin{equation} 
\label{fanomodel} 
G=G_0 
\frac{2\Gamma^2\epsilon^2 \sin^2\phi}
{z^2 t_d^4 + 2\Gamma^2\epsilon^2 - z^2 t_d^4 \cos(2\phi) 
- 2 z t_d^2 \epsilon \Gamma \sin(2\phi)}. 
\end{equation} 
 
Results of the NRG calculation are compared to the prediction from
Eq.~\ref{fanomodel} in the inset of Fig.~\ref{figa10}g.  We see that general
features are adequately described, but there are subtle differences
due to non-perturbative electron correlation effects.  Numerically
calculated Fano resonance is wider than the semi-analytical prediction
and $G/G_0$ does not drop to zero.  In particular, from
Eq.~\ref{fanomodel} it follows that $G=0$ at $\delta=U/2$ 
($\epsilon=0$) and $G=G_0$ at $\delta=U/2+t_d^2
\tan\phi/\Gamma$.  This is not corroborated by NRG results, 
which show maximal conductance at $\delta=U/2$.

We now return to the description of the results presented in Fig.~\ref{figa10}a
in the regime where $\delta<U/2$. As $\delta$  further decreases, the
system enters a regime of the two-stage Kondo effect 
\cite{corn}. This region
is defined by $J_{\mathrm{eff}}<T_K$ (see also Fig.~\ref{figa10}h),
where $T_K$ is the Kondo temperature, approximately given by the
single quantum dot Kondo temperature, Eq.~\ref{tk} with 
\begin{equation}
\rho_0 J =\frac{2\Gamma}{\pi}\left( \frac{1}{\vert \delta-U/2 \vert} + 
\frac{1}{\vert \delta+U/2 \vert} \right).  
\end{equation}
Just below $\delta<U/2$,  $T$ falls in the interval, given
by $T_K^0 \ll T \ll T_K$, where 
\begin{equation}
T_K^0 \sim T_K \exp(-\alpha T_K/J_{\mathrm{eff}})
\end{equation}
denotes the lower Kondo temperature, corresponding to the gap in the spectral density
$\rho_d(\omega)$ at $\omega=0$ and $\alpha$ is of the order of one~\cite{corn}.
Note that NRG values of the gap in $\rho_d(\omega)$ (open circles and
squares), calculated at $T\ll T_K^0$, follow analytical results for
$T_K^0(\delta)$ when $J_{\mathrm{eff}}<T_K$, see Fig.~\ref{figa10}h, while in
the opposite regime, {\it i.e.} for $J_{\mathrm{eff}}>T_K$, they
approach $J_{\mathrm{eff}}$.

As shown in Fig.~\ref{figa10}a for $0.3D \lsim \delta <U/2$, $G/G_0$
calculated at $T=10^{-9}D$ follows results obtained in the single 
quantum dot case and approaches value 1. 
The spin quantum number $S$ in
Fig.~\ref{figa10}b reaches the value $S\sim 0.8$, consistent with the
result obtained for a system of two decoupled spin-1/2 particles, where
$\langle {\bf \hat S}^2\rangle=3/2$. This result is also in agreement
with $\langle n\rangle\sim 2$ and the small value of the spin-spin
correlation function $\langle {\bf S}_a \cdot {\bf S}_d \rangle$, presented
in Fig.~\ref{figa10}c and \ref{figa10}e respectively.

With further decreasing of $\delta$, $G/G_0$ suddenly drops to zero at
$\delta\lsim 0.3D$. This sudden drop is approximately given by $T\lsim
T_K^0(\delta)$, see Figs.~\ref{figa10}a and h. At this point the Kondo
hole opens in $\rho_d(\omega)$ at $\omega=0$, which in turn leads to a
drop in the conductivity. The position of this sudden drop in terms of
$\delta$ is rather insensitive to the chosen $T$, as apparent from
Fig.~\ref{figa10}h.

Below $\delta\lsim 0.25D$, which corresponds to the condition
$J_{\mathrm{eff}}\sim T_K(\delta)$, also presented in
Fig.~\ref{figa10}h, the system crosses over from the two stage Kondo
regime to a regime where spins on DQD form a singlet. In this case $S$
decreases and $\langle {\bf S}_a \cdot {\bf S}_d \rangle$ shows strong
antiferromagnetic correlations, Figs.~\ref{figa10}b and e.  The lowest
energy scale in the system is $J_{\mathrm{eff}}$, which is supported by
the observation that the size of the gap in $\rho_d(\omega)$ (open
circles in Figs.~\ref{figa10}h) is approximately given by $J_{\mathrm{eff}}$. 
The
main difference between $t_d/D=0.001$ and $t_d/D=0.01$ comes from
different values of $J_{\mathrm{eff}}=4t_d^2/U$. Since in the latter
case $J_{\mathrm{eff}}$ is larger, the system enters the AFM singlet
regime at much larger values of $\delta$, as can be seen from
comparison of Figs.~\ref{figa10}g and f. Consequently, the regime of
enhanced conductance shrinks.

\subsection{Effect of the on-site-energy splitting}

We again explore the effect of unequal on-site energies. Unlike
in section \ref{splitting1}, we now introduce the asymmetry
so that only the level $a$ is shifted by parameter $\kappa$:
$\delta_d=\delta$ and $\delta_a=\delta+\kappa$. We are mainly
interested in the parameter range where for $\kappa=0$ 
the system exhibits the two-stage Kondo effect.

This study will be conducted by computing the impurity susceptibility
\begin{equation}
\chi_{\mathrm{imp}}(T)=\frac{\ok{g\mu_B}^2}{k_B T} \ok{ \langle S_z^2 \rangle
- \langle S_z^2 \rangle_0 },
\end{equation}
where the first expectation value refers to the system with the
double quantum dot, while the second refers to the system without
the dots. In traces of $T \chi_{\mathrm{imp}}$, the two-stage
Kondo effect manifests as two successive decreases of the susceptibility,
at $T \sim T_K$ from around $0.5$ to $0.25$ and at $T \sim T_K^0$
from around $0.25$ to $0$ \cite{corn}. We now investigate how this behavior 
changes when $\kappa \neq 0$ by comparing temperature
dependent susceptibilities calculated for a range of parameters $\kappa$.

\begin{figure}[htbp]
\includegraphics[height=10cm,angle=-90]{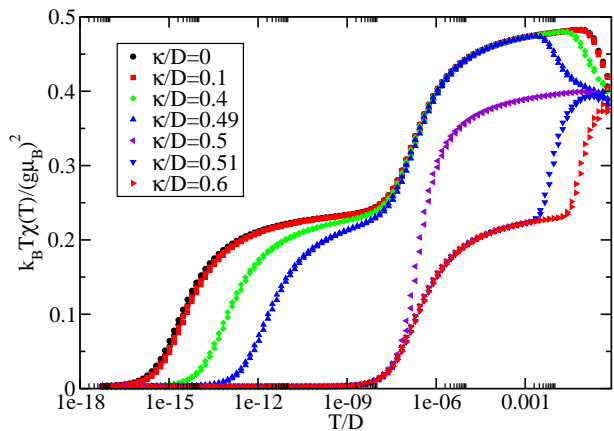}
\caption{Dynamic magnetic susceptibility $k_B T \chi(T)/(g\mu_B)^2$
for different $\kappa$. 
Parameters are $U/D=1$, $\Gamma/D=0.03$, $t_d/D=8\times 10^{-5}$
and $\delta/D=0$. These calculations were performed using
a different NRG discretization scheme \cite{campo2005}
with $\Lambda=4$ and keeping 600 states.
}
\label{figkappa}
\end{figure}

We find that the two stage Kondo effect is robust 
against $\kappa \neq 0$ perturbation. 
In fact, it survives until $\kappa=U/2$, see Fig.~\ref{figkappa}.
For $\kappa<U/2$ the main effect of $\kappa$ is to reduce the 
lower Kondo temperature $T_K^0$. For $\kappa > U/2$, 
dot $a$ becomes irrelevant at low temperatures and dot $d$
is in the usual Kondo regime.

The transition from $\kappa<U/2$ to $\kappa>U/2$ is continuous despite
a rather abrupt change in $\chi_{imp}$ caused by a rapid change in
occupation of dot $a$ that occurs with increasing $\kappa$ on an
energy scale proportional to $t_d^2$.
For $\kappa=U/2$, $\epsilon_a=\delta_a-U/2=0$ and the side-coupled
dot is in the valence fluctuation regime.
For two uncoupled dots, one in local-moment regime and the
other in valence-fluctuation regime, the magnetic moment  
$k_B T \chi/(g\mu_B)^2$ is expected to be $1/4+1/6=5/12$. At high temperatures,
$\chi$ is indeed in the vicinity of this value. 

\section{Conclusions}

In this paper we have explored different regimes of the side-coupled
DQD. 
When quantum dots
are {\it strongly coupled}, wide regions of enhanced, nearly unitary
conductance exist due to the underlying Kondo physics.  Analytical
estimates for their positions, widths, as well as for the
corresponding Kondo temperatures, are given and numerically
verified. When two electrons occupy DQD, conductance is zero due to
formation of the spin-singlet state which is effectively decoupled
from the leads. In this regime most physical properties of the coupled 
DQD can be predicted using the eigenvalue diagram of the isolated
DQD system. This happens because energy separations of different DQD
states are much larger than the Kondo temperature.  Introducing the
on-site-energy splitting between DQD induces new, even though not
unexpected effects within the Kondo regime. With increasing
level-splitting, the Kondo plateau is expanded when the level of the
embedded dot is kept near the local particle-hole limit. In the other
case, when, in contrast, the level of the side-coupled dot is kept
near the particle-hole limit, Kondo plateau splits at finite $T$ due
to abrupt decrease of $T_K$ with increasing level-splitting.

When quantum dots are {\it weakly coupled} Fano resonance
appears in the valence fluctuation regime. Its width is enhanced as a
consequence of interactions which should facilitate experimental
observation. Unitary conductance exists when two electrons occupy DQD
due to a two-stage Kondo effect as long as the temperature of the
system is well below $T_K$ and above $T_K^0$.  The experimental
signature of the two-stage Kondo effect in weakly coupled regime
should materialize through the inter-dot-coupling sensitive width of
the enhanced conductance vs. gate voltage. Experimental observation of
the two-stage Kondo effect would require no external magnetic field as
used in Ref. \cite{two_stage}.  The inter-dot coupling strength can be
experimentally varied using gate electrodes
\cite{dqd-expr1,dqd-expr2,dqd-expr3,pcdqd}.

\begin{acknowledgments} 
Authors acknowledge useful discussions with A. Ram\v sak,
C.D. Batista, S.A. Trugman, and the financial support of the SRA under
grant P1-0044.
\end{acknowledgments}

\appendix*

\section{Density-matrix NRG in (Q,S) subspace basis}

Density-matrix numerical renormalization group (DM-NRG) \cite{hofstetter} 
was originally implemented for a NRG iteration that uses subspaces 
with well defined charge $Q$ and spin projection $S_z$ and it was used to 
determine the effect of magnetic field on spin-projected spectral densities.
In the absence of magnetic field, when rotational invariance in the spin
space is recovered, it is advantageous to use subspaces
with well defined charge $Q$ and spin $S$, which leads to greatly
enhanced numerical efficiency.

Here we show how DM-NRG can be implemented in (Q,S) subspace basis
in the case of a single conduction channel.

Density-matrix at the last iteration step (estimated using the truncated
basis) is:
\begin{equation}
\rho=\frac{1}{Z} \sum_{QSS_z\omega} \exp(-\beta E_{QS\omega})
\ket{QSS_z\omega}\bra{QSS_z\omega},
\end{equation}
where $\omega$ enumerates different states in each $(Q,S)$ subspace and
the grand-canonical statistical sum $Z$ is
\begin{equation}
Z=\tr\left[\exp(-\beta H)\right]
=\sum_{QSS_z\omega} \exp(-\beta E_{QS\omega}).
\end{equation}

Unitary transformation of states from $N$ to $N+1$-th stage is
\begin{equation}
\ket{QSS_z\omega}_{N+1}=\sum_{r\alpha} U_{QS}(\omega|r\alpha)
\ket{QSS_zr \alpha}_{N+1},
\end{equation}
where $U_{QS}$ are the corresponding unitary transformation matrices
obtained during the diagonalization step of the NRG
and the $\ket{QSS_z\alpha r}_{N+1}$ states are defined as
\cite{sia1}
\begin{equation}
\begin{split}
\ket{QSS_zr1}_{N+1} &= \ket{Q+1,SS_zr}_N \\
\ket{QSS_zr2}_{N+1} &= 
u f^\dag_{(N+1)\up} \ket{Q,S-\pol,S_z-\spol,r}_N \\
&+ v f^\dag_{(N+1) \dow} \ket{Q,S-\spol,S_z+\spol,r}_N \\
\ket{QSS_zr3}_{N+1} &= w f^\dag_{(N+1)\up} \ket{Q,S+\spol,S_z-\spol,r}_N \\
&+ y f^\dag_{(N+1) \dow} \ket{Q,S+\spol,S_z+\spol,r}_N \\
\ket{QSS_zr4}_{N+1} &= f^\dag_{(N+1)\up} f^\dag_{(N+1)\dow} 
\ket{Q-1,SS_zr}_N \ket{\up\dow},
\end{split}
\end{equation}
where $f^\dag_{(N+1)\sigma}$ is the creation operator for electrons
on the $(N+1)$-th site of the ``hopping Hamiltonian'' and 
$u, v, w, y$ are the Clebsch-Gordan coefficients 
\begin{equation*}
\begin{split}
u &= \ok{\frac{S+S_z}{2S}}^{1/2} \\
v &= \ok{\frac{S-S_z}{2S}}^{1/2} \\
w &= -\ok{\frac{S-S_z+1}{2S+2}}^{1/2} \\
y &= \ok{\frac{S+S_z+1}{2S+2}}^{1/2}
\end{split}
\end{equation*}

The density matrix in the basis of $\ket{QSS_z\alpha r}_{N+1}$ states is 
\begin{equation}
\begin{split}
\rho &= \sum_{QSS_z\omega} \exp(-\beta E_{QS\omega})
\sum_{r \alpha,r' \alpha'}
U_{QS}(\omega|r\alpha) U_{QS}(\omega|r'\alpha') \\
& \ket{QSS_zr\alpha}\bra{QSS_zr'\alpha'}.
\end{split}
\end{equation}

\begin{widetext}

We now perform a partial trace over the states on the additional 
$(N+1)$-th site to obtain projector operators defined on the chain 
of length $N$.
Diagonal projectors (those with $\alpha=\alpha'$) are:
\begin{equation}
\begin{split}
\tr_{N+1} \ok{ \ket{QSS_zr1} \bra{QSS_zr'1} } &= 
\ket{Q+1,SS_zr}_N \bra{Q+1,SS_zr'}_N \\
\tr_{N+1} \ok{ \ket{QSS_zr4} \bra{QSS_zr'4} } &= 
\ket{Q-1,SS_zr}_N \bra{Q-1,SS_zr}_N \\
\tr_{N+1} \ok{ \ket{QSS_zr2} \bra{QSS_zr'2} } &= 
 u^2 \ket{Q,S-\spol,S_z-\spol,r}_N \bra{Q,S-\spol,S_z-\spol,r'}_N\\
&+ v^2 \ket{Q,S-\spol,S_z+\spol,r}_N \bra{Q,S-\spol,S_z+\spol,r'}_N \\
\tr_{N+1} \ok{ \ket{QSS_zr3} \bra{QSS_zr3} } &= 
 w^2 \ket{Q,S+\spol,S_z-\spol,r}_N \bra{Q,S+\spol,S_z-\spol,r'}_N \\
&+ y^2 \ket{Q,S+\spol,S_z+\spol,r}_N \bra{Q,S+\spol,S_z+\spol,r'}_N
\end{split}
\end{equation}

The out-of-diagonal $\alpha=2, \alpha'=3$ terms give zero when summed over
\begin{equation}
\begin{split}
& \sum_{S_z=-S}^S \tr_{N+1}\ket{QSS_zr2}\bra{QSS_zr'3} = \\ 
&= \sum_{S_z=-S+1}^S \ok{
- \sqrt{\frac{(S+S_z)(S-S_z+1)}{2S(2S+2)}} 
+ \sqrt{\frac{(S-S_z+1)(S+S_z)}{2S(2S+2)}} 
} \\
&\quad \ket{Q,S-\spol,S_z-\spol,r}_N \bra{Q,S+\spol,S_z-\spol,r'}_N = 0
\end{split}
\end{equation}
where we took into account that $v_{SS}=0$ and $u_{S,-S}=0$ in order to
expand the range of summation index $S_z$.
Similarly we show that $\alpha=3, \alpha'=2$ terms drop.

The $\alpha=\alpha'=2$ terms can be simplified after summing over $S_z$:
\begin{equation}
\begin{split}
&\sum_{S_z=-S}^S \tr_{N+1} \ok{ \ket{QSS_zr2} \bra{QSS_zr'2} }\\
& = \sum_{S_z=-S+1}^S
\ok{ u_{SS_z}^2+v_{S,S_z-1}^2 }
\ket{Q,S-\spol,S_z-\spol,r}_N \bra{Q,S-\spol,S_z-\spol,r'}_N \\
& = 
\frac{2S+1}{2S} 
\sum_{S_z=-(S-\spol)}^{+(S-\spol)}
\ket{Q,S-\spol,S_z}_N \bra{Q,S-\spol,S_z,r'}_N \\
\end{split}
\end{equation}

\end{widetext}

The spin multiplicity of $(QS)_{N+1}$ space is $2S+1$,
while the spin multiplicity of $(Q,S-\spol)_{N}$ is $2S$.
The factor $(2S+1)/(2S)$ is therefore merely a normalization
factor. In the last line we emphasized that in the $N$-site
space the $S_z$ runs over all permissible values for spin $S-\spol$.

By analogy we show that the $\alpha=\alpha'=3$ terms also simplify:
\begin{equation}
\begin{split}
&\sum_{S_z=-S}^S \tr_{N+1} \ok{ \ket{QSS_zr3} \bra{QSS_zr3} }\\
&= \frac{2S+1}{2S+2}
\sum_{S_z=-(S+\spol)}^{S+\spol}
\ket{Q,S+\spol,S_z,r}_N \bra{Q,S+\spol,S_z,r'}_N
\end{split}
\end{equation}

Again, the spin multiplicity of the $(Q,S+\spol)_N$ space
is $2S+2$ and the factor $(2S+1)/(2S+2)$ takes care of the
correct normalization.

Non-zero partial traces of projector operators are therefore:
\begin{equation}
\begin{split}
&\sum_{S_z} \tr_{N+1} \ket{QSS_zr\alpha} \bra{QSS_zr'\alpha'} \\
&= \delta_{\alpha\alpha'} c_\alpha(S) 
\sum_{S_{z\alpha}}
\ket{Q_\alpha S_\alpha S_{z\alpha} r}_N
\bra{Q_\alpha S_\alpha S_{z\alpha} r'}_N
\end{split}
\end{equation}
with $c_1=c_4=1$, $c_2=\frac{2S+1}{2S}$, $c_3=\frac{2S+1}{2S+2}$,
$Q_1=Q+1$, $Q_2=Q_3=Q$, $Q_4=Q-1$,
$S_1=S_4=S$, $S_2=S-\spol$, $S_3=S+\spol$
and corresponding $S_{z\alpha}$ ranges over all possible values
for a given $S_\alpha$.
The reduced density matrix remains diagonal in its $(QS)$ subspace
index. 

In general we therefore have
\begin{equation}
\begin{split}
&\rho^{N+1}_\mathrm{reduced} = \sum_{QSS_z} \sum_{\omega\omega'} C^{QS,N+1}_{\omega\omega'}
\ket{QSS_z\omega}\bra{QSS_z\omega'} \\
&= \sum_{QS} \sum_{\omega\omega'} C^{QS,N+1}_{\omega\omega'}
\sum_{r\alpha,r'}
U_{QS}(\omega|r\alpha)\\
& U_{QS}(\omega'|r'\alpha)
c_\alpha(S) 
 \sum_{S_{z\alpha}} 
\ket{Q_\alpha S_\alpha S_{z\alpha} r}_N
\bra{Q_\alpha S_\alpha S_{z\alpha} r'}_N.\\
\end{split}
\end{equation}

This is to be comapred with
\begin{equation}
\rho^{N}_\mathrm{reduced}=\sum_{QSS_z} \sum_{rr'} C^{QS,N}_{rr'}
\ket{QSS_zr}\bra{QSS_zr'}.
\end{equation}

We finally obtain the recursion relation for calculation of coefficients 
$C^{QS,N}_{rr'}$ in the reduced density matrix:

\begin{eqnarray}
\label{recur}
C^{QS,N}_{rr'} &=&
\sum_{\omega\omega'} C^{Q-1,S,N+1}_{\omega\omega'}
U_{Q-1,S}(\omega|r1) U_{Q-1,S}(\omega'|r'1) \nonumber\\
&+&
\sum_{\omega\omega'} C^{Q+1,S,N+1}_{\omega\omega'}
U_{Q+1,S}(\omega|r4) U_{Q+1,S}(\omega'|r'4) \nonumber\\
&+&
\frac{2S+2}{2S+1}\sum_{\omega\omega'} 
C^{Q,S+\spol,N+1}_{\omega\omega'} \nonumber\\
&&\quad U_{Q,S+\spol}(\omega|r2) U_{Q,S+\spol}(\omega'|r'2) \nonumber\\
&+&
\frac{2S}{2S+1} \sum_{\omega\omega'} 
C^{Q,S-\spol,N+1}_{\omega\omega'} \nonumber\\
&&\quad U_{Q,S-\spol}(\omega|r3) U_{Q,S-\spol}(\omega'|r'3)
\end{eqnarray}

This is the main result of the derivation. Using known
$U_{Q,S}$ matrices, recursion in Eq.~\eqref{recur}
is applied after the first NRG run to calculate 
reduced density matrices for all chain lengths.
In another NRG run, the spectral density functions are then
calculated with respect to the reduced density matrices:
\begin{equation}
\begin{split}
A^N_{d\sigma}(\omega) 
&= \sum_{ijm} 
\Bigl(\braket{j}{d^\dag_\sigma|m}
\braket{j}{d^\dag_\sigma|i} \rho^\mathrm{reduced}_{im}
 \\
&+
\braket{i}{d^\dag_\sigma|m}
\braket{j}{d^\dag_\sigma|m} \rho^\mathrm{reduced}_{ji}\Bigr)
\delta\left(\omega-(E_j-E_m)\right)
\end{split}
\end{equation}

\bibliography{vsi} 

\begin{thebibliography}{29}
\expandafter\ifx\csname natexlab\endcsname\relax\def\natexlab#1{#1}\fi
\expandafter\ifx\csname bibnamefont\endcsname\relax
  \def\bibnamefont#1{#1}\fi
\expandafter\ifx\csname bibfnamefont\endcsname\relax
  \def\bibfnamefont#1{#1}\fi
\expandafter\ifx\csname citenamefont\endcsname\relax
  \def\citenamefont#1{#1}\fi
\expandafter\ifx\csname url\endcsname\relax
  \def\url#1{\texttt{#1}}\fi
\expandafter\ifx\csname urlprefix\endcsname\relax\def\urlprefix{URL }\fi
\providecommand{\bibinfo}[2]{#2}
\providecommand{\eprint}[2][]{\url{#2}}

\bibitem[{\citenamefont{Jeong et~al.}(2001)\citenamefont{Jeong, Chang, and
  Melloch}}]{dqd-expr1}
\bibinfo{author}{\bibfnamefont{H.}~\bibnamefont{Jeong}},
  \bibinfo{author}{\bibfnamefont{A.~M.} \bibnamefont{Chang}}, \bibnamefont{and}
  \bibinfo{author}{\bibfnamefont{M.~R.} \bibnamefont{Melloch}},
  \bibinfo{journal}{Science} \textbf{\bibinfo{volume}{293}},
  \bibinfo{pages}{2221} (\bibinfo{year}{2001}).

\bibitem[{\citenamefont{Craig et~al.}(2004)\citenamefont{Craig, Taylor, Lester,
  Marcus, Hanson, and Gossard}}]{dqd-expr2}
\bibinfo{author}{\bibfnamefont{N.~J.} \bibnamefont{Craig}},
  \bibinfo{author}{\bibfnamefont{J.~M.} \bibnamefont{Taylor}},
  \bibinfo{author}{\bibfnamefont{E.~A.} \bibnamefont{Lester}},
  \bibinfo{author}{\bibfnamefont{C.~M.} \bibnamefont{Marcus}},
  \bibinfo{author}{\bibfnamefont{M.~P.} \bibnamefont{Hanson}},
  \bibnamefont{and} \bibinfo{author}{\bibfnamefont{A.~C.}
  \bibnamefont{Gossard}}, \bibinfo{journal}{Science}
  \textbf{\bibinfo{volume}{304}}, \bibinfo{pages}{565} (\bibinfo{year}{2004}).

\bibitem[{\citenamefont{Holleitner et~al.}(2002)\citenamefont{Holleitner,
  Blick, H{\:u}ttel, Eberl, and Kotthaus}}]{dqd-expr3}
\bibinfo{author}{\bibfnamefont{A.~W.} \bibnamefont{Holleitner}},
  \bibinfo{author}{\bibfnamefont{R.~H.} \bibnamefont{Blick}},
  \bibinfo{author}{\bibfnamefont{A.~K.} \bibnamefont{H{\:u}ttel}},
  \bibinfo{author}{\bibfnamefont{K.}~\bibnamefont{Eberl}}, \bibnamefont{and}
  \bibinfo{author}{\bibfnamefont{J.~P.} \bibnamefont{Kotthaus}},
  \bibinfo{journal}{Science} \textbf{\bibinfo{volume}{297}},
  \bibinfo{pages}{70} (\bibinfo{year}{2002}).

\bibitem[{\citenamefont{Chen et~al.}(2004)\citenamefont{Chen, Chang, and
  Melloch}}]{pcdqd}
\bibinfo{author}{\bibfnamefont{J.~C.} \bibnamefont{Chen}},
  \bibinfo{author}{\bibfnamefont{A.~M.} \bibnamefont{Chang}}, \bibnamefont{and}
  \bibinfo{author}{\bibfnamefont{M.~R.} \bibnamefont{Melloch}},
  \bibinfo{journal}{Phys. Rev. Lett.} \textbf{\bibinfo{volume}{92}},
  \bibinfo{pages}{176801} (\bibinfo{year}{2004}).

\bibitem[{\citenamefont{Hofstetter and Schoeller}(2003)}]{qptmultilevel}
\bibinfo{author}{\bibfnamefont{W.}~\bibnamefont{Hofstetter}} \bibnamefont{and}
  \bibinfo{author}{\bibfnamefont{H.}~\bibnamefont{Schoeller}},
  \bibinfo{journal}{Phys. Rev. Lett.} \textbf{\bibinfo{volume}{88}},
  \bibinfo{pages}{016803} (\bibinfo{year}{2003}).

\bibitem[{\citenamefont{Hofstetter and Zarand}(2004)}]{hofstetter2004}
\bibinfo{author}{\bibfnamefont{W.}~\bibnamefont{Hofstetter}} \bibnamefont{and}
  \bibinfo{author}{\bibfnamefont{G.}~\bibnamefont{Zarand}},
  \bibinfo{journal}{Phys. Rev. B} \textbf{\bibinfo{volume}{69}},
  \bibinfo{pages}{235301} (\bibinfo{year}{2004}).

\bibitem[{\citenamefont{van~der Wiel et~al.}(2002)\citenamefont{van~der Wiel,
  Franceschi, Elzerman, Tarucha, Kouwenhoven, Motohisa, Nakajima, and
  Fukui}}]{two_stage}
\bibinfo{author}{\bibfnamefont{W.~G.} \bibnamefont{van~der Wiel}},
  \bibinfo{author}{\bibfnamefont{S.~D.} \bibnamefont{Franceschi}},
  \bibinfo{author}{\bibfnamefont{J.~M.} \bibnamefont{Elzerman}},
  \bibinfo{author}{\bibfnamefont{S.}~\bibnamefont{Tarucha}},
  \bibinfo{author}{\bibfnamefont{L.~P.} \bibnamefont{Kouwenhoven}},
  \bibinfo{author}{\bibfnamefont{J.}~\bibnamefont{Motohisa}},
  \bibinfo{author}{\bibfnamefont{F.}~\bibnamefont{Nakajima}}, \bibnamefont{and}
  \bibinfo{author}{\bibfnamefont{T.}~\bibnamefont{Fukui}},
  \bibinfo{journal}{Phys. Rev. Lett.} \textbf{\bibinfo{volume}{88}},
  \bibinfo{pages}{126803} (\bibinfo{year}{2002}).

\bibitem[{\citenamefont{Granger et~al.}(2005)\citenamefont{Granger, Kastner,
  Radu, Hanson, and Gossard}}]{granger2005}
\bibinfo{author}{\bibfnamefont{G.}~\bibnamefont{Granger}},
  \bibinfo{author}{\bibfnamefont{M.~A.} \bibnamefont{Kastner}},
  \bibinfo{author}{\bibfnamefont{I.}~\bibnamefont{Radu}},
  \bibinfo{author}{\bibfnamefont{M.~P.} \bibnamefont{Hanson}},
  \bibnamefont{and} \bibinfo{author}{\bibfnamefont{A.~C.}
  \bibnamefont{Gossard}}, \bibinfo{journal}{Phys. Rev. B}
  \textbf{\bibinfo{volume}{72}}, \bibinfo{pages}{165309}
  (\bibinfo{year}{2005}).

\bibitem[{\citenamefont{Kobayashi et~al.}(2002)\citenamefont{Kobayashi, Aikawa,
  Katsumoto, and Iye}}]{ringfano}
\bibinfo{author}{\bibfnamefont{K.}~\bibnamefont{Kobayashi}},
  \bibinfo{author}{\bibfnamefont{H.}~\bibnamefont{Aikawa}},
  \bibinfo{author}{\bibfnamefont{S.}~\bibnamefont{Katsumoto}},
  \bibnamefont{and} \bibinfo{author}{\bibfnamefont{Y.}~\bibnamefont{Iye}},
  \bibinfo{journal}{Phys. Rev. Lett.} \textbf{\bibinfo{volume}{88}},
  \bibinfo{pages}{256806} (\bibinfo{year}{2002}).

\bibitem[{\citenamefont{Kobayashi et~al.}(2004)\citenamefont{Kobayashi, Aikawa,
  Sano, Katsumoto, and Iye}}]{scfano}
\bibinfo{author}{\bibfnamefont{K.}~\bibnamefont{Kobayashi}},
  \bibinfo{author}{\bibfnamefont{H.}~\bibnamefont{Aikawa}},
  \bibinfo{author}{\bibfnamefont{A.}~\bibnamefont{Sano}},
  \bibinfo{author}{\bibfnamefont{S.}~\bibnamefont{Katsumoto}},
  \bibnamefont{and} \bibinfo{author}{\bibfnamefont{Y.}~\bibnamefont{Iye}},
  \bibinfo{journal}{Phys. Rev. B} \textbf{\bibinfo{volume}{70}},
  \bibinfo{pages}{035319} (\bibinfo{year}{2004}).

\bibitem[{\citenamefont{Bulka and Stefanski}(2001)}]{bulka1}
\bibinfo{author}{\bibfnamefont{B.~R.} \bibnamefont{Bulka}} \bibnamefont{and}
  \bibinfo{author}{\bibfnamefont{P.}~\bibnamefont{Stefanski}},
  \bibinfo{journal}{Phys. Rev. Lett.} \textbf{\bibinfo{volume}{86}},
  \bibinfo{pages}{5128} (\bibinfo{year}{2001}).

\bibitem[{\citenamefont{Stefanski et~al.}(2004)\citenamefont{Stefanski,
  Tagliacozzo, and Bulka}}]{bulka2}
\bibinfo{author}{\bibfnamefont{P.}~\bibnamefont{Stefanski}},
  \bibinfo{author}{\bibfnamefont{A.}~\bibnamefont{Tagliacozzo}},
  \bibnamefont{and} \bibinfo{author}{\bibfnamefont{B.~R.} \bibnamefont{Bulka}},
  \bibinfo{journal}{Phys. Rev. Lett.} \textbf{\bibinfo{volume}{93}},
  \bibinfo{pages}{186805} (\bibinfo{year}{2004}).

\bibitem[{\citenamefont{Lara et~al.}()\citenamefont{Lara, Orellana, Yanez, and
  Anda}}]{sidedouble}
\bibinfo{author}{\bibfnamefont{G.~A.} \bibnamefont{Lara}},
  \bibinfo{author}{\bibfnamefont{P.~A.} \bibnamefont{Orellana}},
  \bibinfo{author}{\bibfnamefont{J.~M.} \bibnamefont{Yanez}}, \bibnamefont{and}
  \bibinfo{author}{\bibfnamefont{E.~V.} \bibnamefont{Anda}},
  \bibinfo{note}{cond-mat/0411661}.

\bibitem[{\citenamefont{Kim and Hershfield}(2001)}]{suppression}
\bibinfo{author}{\bibfnamefont{T.-S.} \bibnamefont{Kim}} \bibnamefont{and}
  \bibinfo{author}{\bibfnamefont{S.}~\bibnamefont{Hershfield}},
  \bibinfo{journal}{Phys. Rev. B} \textbf{\bibinfo{volume}{63}},
  \bibinfo{pages}{245326} (\bibinfo{year}{2001}).

\bibitem[{\citenamefont{Apel et~al.}(2004)\citenamefont{Apel, Davidovich, Anda,
  Chiappe, and Busser}}]{topology}
\bibinfo{author}{\bibfnamefont{V.~M.} \bibnamefont{Apel}},
  \bibinfo{author}{\bibfnamefont{M.~A.} \bibnamefont{Davidovich}},
  \bibinfo{author}{\bibfnamefont{E.~V.} \bibnamefont{Anda}},
  \bibinfo{author}{\bibfnamefont{G.}~\bibnamefont{Chiappe}}, \bibnamefont{and}
  \bibinfo{author}{\bibfnamefont{C.~A.} \bibnamefont{Busser}},
  \bibinfo{journal}{Eur. Phys. J. B} \textbf{\bibinfo{volume}{40}},
  \bibinfo{pages}{365} (\bibinfo{year}{2004}).

\bibitem[{\citenamefont{Kang et~al.}(2001)\citenamefont{Kang, Cho, Kim, and
  Shin}}]{kang}
\bibinfo{author}{\bibfnamefont{K.}~\bibnamefont{Kang}},
  \bibinfo{author}{\bibfnamefont{S.~Y.} \bibnamefont{Cho}},
  \bibinfo{author}{\bibfnamefont{J.-J.} \bibnamefont{Kim}}, \bibnamefont{and}
  \bibinfo{author}{\bibfnamefont{S.-C.} \bibnamefont{Shin}},
  \bibinfo{journal}{Phys. Rev. B} \textbf{\bibinfo{volume}{63}},
  \bibinfo{pages}{113304} (\bibinfo{year}{2001}).

\bibitem[{\citenamefont{Cornaglia and Grempel}(2005)}]{corn}
\bibinfo{author}{\bibfnamefont{P.~S.} \bibnamefont{Cornaglia}}
  \bibnamefont{and} \bibinfo{author}{\bibfnamefont{D.~R.}
  \bibnamefont{Grempel}}, \bibinfo{journal}{Phys. Rev. B}
  \textbf{\bibinfo{volume}{71}}, \bibinfo{pages}{075305}
  (\bibinfo{year}{2005}).

\bibitem[{\citenamefont{Vojta et~al.}(2002)\citenamefont{Vojta, Bulla, and
  Hofstetter}}]{vojta}
\bibinfo{author}{\bibfnamefont{M.}~\bibnamefont{Vojta}},
  \bibinfo{author}{\bibfnamefont{R.}~\bibnamefont{Bulla}}, \bibnamefont{and}
  \bibinfo{author}{\bibfnamefont{W.}~\bibnamefont{Hofstetter}},
  \bibinfo{journal}{Phys. Rev. B} \textbf{\bibinfo{volume}{65}},
  \bibinfo{pages}{140405} (\bibinfo{year}{2002}).

\bibitem[{\citenamefont{Glazman and Raikh}(1988)}]{glazmanraikh}
\bibinfo{author}{\bibfnamefont{L.~I.} \bibnamefont{Glazman}} \bibnamefont{and}
  \bibinfo{author}{\bibfnamefont{M.~E.} \bibnamefont{Raikh}},
  \bibinfo{journal}{JETP Lett.} \textbf{\bibinfo{volume}{47}},
  \bibinfo{pages}{452} (\bibinfo{year}{1988}).

\bibitem[{\citenamefont{Meir and Wingreen}(1992)}]{meirwingreen}
\bibinfo{author}{\bibfnamefont{Y.}~\bibnamefont{Meir}} \bibnamefont{and}
  \bibinfo{author}{\bibfnamefont{N.~S.} \bibnamefont{Wingreen}},
  \bibinfo{journal}{Phys. Rev. Lett.} \textbf{\bibinfo{volume}{68}},
  \bibinfo{pages}{2512} (\bibinfo{year}{1992}).

\bibitem[{\citenamefont{Wilson}(1975)}]{wilson}
\bibinfo{author}{\bibfnamefont{K.~G.} \bibnamefont{Wilson}},
  \bibinfo{journal}{Rev. Mod. Phys.} \textbf{\bibinfo{volume}{47}},
  \bibinfo{pages}{773} (\bibinfo{year}{1975}).

\bibitem[{\citenamefont{Costi}(2001)}]{magnetocosti}
\bibinfo{author}{\bibfnamefont{T.~A.} \bibnamefont{Costi}},
  \bibinfo{journal}{Phys. Rev. B} \textbf{\bibinfo{volume}{64}},
  \bibinfo{pages}{241310} (\bibinfo{year}{2001}).

\bibitem[{\citenamefont{Krishna-Murthy
  et~al.}(1980)\citenamefont{Krishna-Murthy, Wilkins, and Wilson}}]{sia1}
\bibinfo{author}{\bibfnamefont{H.~R.} \bibnamefont{Krishna-Murthy}},
  \bibinfo{author}{\bibfnamefont{J.~W.} \bibnamefont{Wilkins}},
  \bibnamefont{and} \bibinfo{author}{\bibfnamefont{K.~G.}
  \bibnamefont{Wilson}}, \bibinfo{journal}{Phys. Rev. B}
  \textbf{\bibinfo{volume}{21}}, \bibinfo{pages}{1003} (\bibinfo{year}{1980}).

\bibitem[{\citenamefont{Hofstetter}(2000)}]{hofstetter}
\bibinfo{author}{\bibfnamefont{W.}~\bibnamefont{Hofstetter}},
  \bibinfo{journal}{Phys. Rev. Lett.} \textbf{\bibinfo{volume}{85}},
  \bibinfo{pages}{1508} (\bibinfo{year}{2000}).

\bibitem[{\citenamefont{Hewson}(1993)}]{hewson}
\bibinfo{author}{\bibfnamefont{A.~C.} \bibnamefont{Hewson}},
  \emph{\bibinfo{title}{The Kondo problem to heavy fermions}}
  (\bibinfo{publisher}{Cambridge University Press}, \bibinfo{year}{1993}).

\bibitem[{\citenamefont{Haldane}(1978)}]{hald}
\bibinfo{author}{\bibfnamefont{F.}~\bibnamefont{Haldane}}, \bibinfo{journal}{J.
  Phys. C: Solid State Phys.} \textbf{\bibinfo{volume}{11}},
  \bibinfo{pages}{5015} (\bibinfo{year}{1978}).

\bibitem[{\citenamefont{Costi et~al.}(1994)\citenamefont{Costi, Hewson, and
  Zlatic}}]{costi}
\bibinfo{author}{\bibfnamefont{T.~A.} \bibnamefont{Costi}},
  \bibinfo{author}{\bibfnamefont{A.~C.} \bibnamefont{Hewson}},
  \bibnamefont{and} \bibinfo{author}{\bibfnamefont{V.}~\bibnamefont{Zlatic}},
  \bibinfo{journal}{J. Phys.: Condens. Matter} \textbf{\bibinfo{volume}{6}},
  \bibinfo{pages}{2519} (\bibinfo{year}{1994}).

\bibitem[{\citenamefont{Langreth}(1966)}]{langreth66}
\bibinfo{author}{\bibfnamefont{D.~C.} \bibnamefont{Langreth}},
  \bibinfo{journal}{Phys. Rev.} \textbf{\bibinfo{volume}{150}},
  \bibinfo{pages}{516} (\bibinfo{year}{1966}).

\bibitem[{\citenamefont{Campo and Oliveira}(2005)}]{campo2005}
\bibinfo{author}{\bibfnamefont{V.~L.} \bibnamefont{Campo}} \bibnamefont{and}
  \bibinfo{author}{\bibfnamefont{L.~N.} \bibnamefont{Oliveira}},
  \bibinfo{journal}{Phys. Rev. B} \textbf{\bibinfo{volume}{72}},
  \bibinfo{pages}{104432} (\bibinfo{year}{2005}).

\end{thebibliography}
 
\end{document}